\documentclass[11pt]{article}

\makeatletter
\def\maxwidth{ %
  \ifdim\Gin@nat@width>\linewidth
    \linewidth
  \else
    \Gin@nat@width
  \fi
}
\makeatother

\RequirePackage{amsmath,amssymb}
\RequirePackage{amsthm}
\RequirePackage{natbib}
\RequirePackage[colorlinks,allcolors=blue]{hyperref}

\usepackage{fullpage}
\usepackage{setspace}
\usepackage{nicefrac}

\usepackage{bbm,graphicx,bm} 
\usepackage{enumitem}
\usepackage{multirow}
\usepackage[all,import]{xy}
\usepackage{appendix}
\usepackage{comment}
\usepackage{color}
\usepackage{soul}
\usepackage{pdflscape}
\usepackage{subcaption}

\usepackage{booktabs}
\usepackage{longtable}
\usepackage{array}
\usepackage[table]{xcolor}
\usepackage{wrapfig}
\usepackage{float}
\usepackage{colortbl}
\usepackage{pdflscape}
\usepackage{tabu}
\usepackage{threeparttable}
\usepackage{threeparttablex}
\usepackage[normalem]{ulem}
\usepackage{makecell}
\usepackage{afterpage}

\usepackage{flafter} 

\theoremstyle{definition}

\newcommand{\E}{\ensuremath{\mathbb{E}}}

\newcommand{\calD}{\ensuremath{\mathcal{D}}}

\def\b1{\boldsymbol{1}}

\definecolor{RED}{RGB}{255,0,0}

\usepackage{soul}
\usepackage[utf8]{inputenc}

\title{Statistical methods to estimate the impact of gun policy on gun violence}

\author{Eli Ben-Michael\thanks{Heinz College of Information Systems \& Public Policy and Department of Statistics \& Data Science, Carnegie Mellon University}, Mitchell L. Doucette\thanks{Center for Gun Violence Solutions, Johns Hopkins Bloomberg School of Public Health}, Avi Feller\thanks{Goldman School of Public Policy \& Department of Statistics, University of California, Berkeley}, Alexander D. McCourt,\thanks{Department of Health Policy and Management and Center for Gun Violence Solutions, Johns Hopkins Bloomberg School of Public Health} and  Elizabeth A. Stuart\thanks{Department of Biostatistics, Department of Mental Health, Department of Health Policy and Management, ad Center for Gun Violence Solutions, Johns Hopkins Bloomberg School of Public Health}}
\date{}

\begin{document}

\maketitle

\thispagestyle{empty}
\pagenumbering{gobble}

\begin{abstract}
\singlespacing
Gun violence is a critical public health and safety concern in the United States. There is considerable variability in policy proposals meant to curb gun violence, ranging from increasing gun availability to deter potential assailants (e.g., concealed carry laws or arming school teachers) to restricting access to firearms (e.g., universal background checks or banning assault weapons). Many studies use state-level variation in the enactment of these policies in order to quantify their effect on gun violence. In this paper, we discuss the policy trial emulation framework for evaluating the impact of these policies, and show how to apply this framework to estimating impacts via difference-in-differences and synthetic controls when there is staggered adoption of policies across jurisdictions, estimating the impacts of right-to-carry laws on violent crime as a case study.

\end{abstract}

\clearpage
\pagenumbering{arabic}
\onehalfspacing

\section{Introduction}

Rigorous evaluations of public policy are of the utmost importance. Understanding whether a policy has had its intended effect---or an unintended one---is essential for making future policy decisions.
This is especially true for firearm policy.
The United States has the highest rate of firearm violence among developed nations, with a gun homicide rate roughly 25\% higher than other high-income countries \citep{bangalore_gun_2013}. This relationship only worsened in the mid 2010's, with gun violence surging in 2020 \citep{schleimer_firearm_2021}. Firearm violence is the leading cause of death for American adolescents under the age of 21 and firearm homicides disproportionately impact  young Black Americans \citep{davis_us_2023}. Over the past several decades, lawmakers have attempted to address gun violence in many ways, but the most prominent approach has been laws that regulate the purchase, possession, and use of guns. 

Understanding the impact of such policies is important but challenging, and researchers must overcome several significant hurdles.
The research has been severely underfunded overall and relative to other leading causes of death \citep{ault_gun_2021, stark_funding_2017}. Data availability has been a barrier for researchers: analyses of gun policies are hindered by severe limitations on the type and scope of data available \citep{rowhani-rahbar_firearm_2019}. There have also been challenges related to statistical methodology: many researchers seek to use variation in state policies to draw causal inferences, but reliable study designs are often difficult to craft. Because states have many different gun laws, it can be challenging to determine how to account for additional policies and which ones to account for. Similarly, though state-level policy variation creates research opportunities, there are only 51 jurisdictions (50 states and Washington, D.C.) eligible for inclusion in these analyses, limiting the combinations of treatment and controls states for comparisons.

Despite these challenges, gun policy studies are quite impactful. Policy makers and advocates use these studies to push for adoption of new, evidence-based gun policies. For these reasons, it is essential that gun policy researchers use rigorous methods to estimate the effects of these laws. In this paper we describe an approach known as policy trial emulation, which uses a design-focused approach to analyze data on states with and without policies of interest over time.  This approach aims to take advantage of longitudinal state-level data to compare outcomes in states that do and do not implement the policy, while adjusting for baseline differences between states in the pre-policy period to disentangle the policy's effects from other differences between states.
This paper highlights the key features of the design and analysis of such evaluations along with associated opportunities and challenges, especially for studying firearm policies in particular.
The goal is to help researchers construct trustworthy studies and to help consumers of research understand which results are trustworthy.

\subsection{Firearm policies and impact evaluations}

Most gun laws in the United States are enacted and implemented at the state level. Though the federal government has largely been inactive, this is also a function of governmental design. The U.S. Constitution enumerates specific powers for the federal government while the Tenth Amendment affords states all other powers not prohibited by it.
This includes ``police power,'' which allows states to regulate for the sake of public health, public safety, and the general welfare. The breadth of these state powers has historically meant states have approached gun policy in varied ways. The variation results
in a patchwork of different regulations. This ``laboratory of democracy'' creates an environment that is ripe for causal inference --- at least in principle.

State gun laws typically seek to regulate who can own or acquire guns, including the purchase process, possession, or use of a gun, or other behaviors related to ownership of guns. State laws typically list categories of individuals who are prohibited from possessing guns (e.g., those convicted of felonies, those with a specific history of alcohol misuse, those convicted of certain juvenile offenses, etc.) \citep{smart_science_2023}. Laws regulating the purchase process typically involve a background check that seeks to determine whether a prospective purchaser is in one of these prohibited categories \citep{vernick_background_2017}. Though many states do not require a background check for private transfers (those not involving a licensed firearms dealer), others require a background check as part of the purchase or acquisition process. These background checks may occur at the point-of-sale or may be implemented as part of a permit or license process \citep{vernick_background_2017, mccourt_purchaser_2020}. 
States have also tried to regulate the use of guns and related behaviors. These policies include laws like Child Access Prevention laws, which require parents to store guns in a specified manner
\citep{webster_association_2004, hepburn_effect_2006, azad_child_2020}.
An alternative direction are ``stand-your-ground'' laws that extend the rights of gun owners to use deadly force for self-defense in public spaces, even in cases where retreating from a conflict would be sufficient \citep{degli_esposti_analysis_2022}. Some states have laws that place limits on the design and manufacture of guns including regulations of private manufacture. Many states have laws that prohibit guns in certain locations, like school grounds, public transit, or government buildings. Among many other laws, states have also traditionally had laws that regulate who can carry a concealed handgun in public.

Researchers have taken advantage of state-level variation in these laws to study the relationship between their enactment and different measures of gun violence. Most of these studies focus on mortality data, with most homicide or suicide data acquired from sources like the National Vital Statistics System. Other studies use crime data, usually acquired from the FBI's historic Uniform Crime Reports or from the newer National Incident-Based Reporting System.
For state laws, researchers either identify and code them through their own legal research or use an existing database \citep[e.g.][]{siegel_firearm-related_2017}. Though many studies are cross-sectional or otherwise descriptive, researchers attempting to make causal inferences have used strategies 
involving policies changing over time.
In particular, these studies generally use what is known as ``panel data,'' consisting of repeated measurements of the outcome of interest over time measured on relevant geographic units (such as states), with individual states adopting or removing the policy of interest at different times.
The chief goal of this paper is to understand how best to evaluate policies with this kind of data.

Studies using panel data have generated a body of scholarship that reveals some trends in the relationship between gun violence and gun policy.
For example, multiple studies of laws requiring a license or permit to purchase a firearm have found reductions in firearm homicide, firearm suicide, mass shooting incidence, and other measures of gun violence associated with enactment of these laws
\citep{mccourt_purchaser_2020,crifasi_effects_2015,crifasi_association_2018,doucette_impact_2021,rubin1974,hasegawa_evaluating_2019,webster_effects_2014,webster_evidence_2020}.
Similar studies on laws requiring a background check at the point of sale (without any license requirement) have failed to find any consistent effect on population-level violence
\citep{mccourt_purchaser_2020,castillo-carniglia_comprehensive_2018, castillo-carniglia_californias_2019,kagawa_repeal_2018, kagawa_effects_2023}. Child Access Prevention Laws have consistently been associated with reductions in suicide, unintentional shootings, and interpersonal violence. Studies of stand-your-ground laws, on the other hand, have regularly found associations with increased violent crime \citep{smart_science_2023}.

\subsection{Case study: the impact of right-to-carry laws on violent crime}

To guide us through the key methodological considerations for estimating the impact of firearm policies, we will focus on a particular case study: measuring the impact of so-called ``right-to-carry'' (RTC) laws.
Many gun policies, including those described above, have been the subject of multiple studies, but perhaps no policy has been examined for as long --- and with such politicization --- as RTC laws.
Academic researchers have been evaluating these laws, otherwise referred to as ``shall-issue'' or ``permitless'' concealed carry weapons (CCW) laws, since the mid 1990s, and as we will discuss below, the policy space has changed significantly over the last 30 years.



RTC laws generally intend to increase the number of citizens who have the right to carry a concealed weapon in public (we defer further policy details to Section~\ref{sec:operationalize}). The changing landscape of state laws governing concealed gun carrying has provided researchers an opportunity to estimate the relationship between changes to these laws and public safety. Early research on CCW laws by \citet{lott_crime_1997}  suggested that RTC law adoption led to reductions in violent crime. The ``more guns, less crime'' hypothesis that was foundational to this work posited that if more civilians were legally carrying firearms in public, criminals would be deterred from engaging in crime as they would not know if a potential target was armed. The study, however, contained serious methodological flaws noted in the literature \citep{webster_flawed_1997,doucette_rightcarry_2019,donohue2019right}. Despite these flaws, this hypothesis and perceived deterrent benefits have been used to argue for lowering restrictions for concealed carry.

More recent research, with access to additional years of data and using newer statistical techniques, has evaluated the impacts of adopting RTC laws on violent crime outcomes \citep[e.g.][]{mcelroy_seemingly_2017,siegel_firearm-related_2017, siegel_impact_2019, crifasi_association_2018, knopov_impact_2019, doucette_rightcarry_2019, doucette_officer-involved_2022, doucette_deregulation_2023, doucette_impact_2023, van_der_wal_marginal_2022, donohue_why_2023, smart_science_2023}.
In this paper, we will take this well-trod path and re-analyze the impact of RTC laws on violent crime (following the data setup in \citet{donohue2019right} in particular) in order to pull out the important steps for gun policy impact evaluations.

\subsection{Plan for the paper}
In this paper, we will go through the important elements of impact evaluations of firearm policies using modern methods. We separate out our development into three key steps. First and foremost, in Section~\ref{sec:operationalize} we discuss the important design considerations that allow us to be clear about \emph{what} effects we are attempting to estimate. Second, in Section~\ref{sec:single} we will begin our discussion of \emph{how} to estimate these effects by focusing on estimating effects for a single unit in comparative case studies.
Third, we turn to estimating overall effects in Section~\ref{sec:aggregate}.
In Section~\ref{sec:challenges} we discuss additional methodological challenges for firearm policy impact evaluations.
Throughout this paper we pair a formalization of the statistical issues with a practical discussion about implementation.

\section{Operationalizing an impact evaluation}
\label{sec:operationalize}
To begin our discussion, we will go over key considerations for operationalizing an observational impact evaluation with panel data. These decisions are independent of the particular estimation strategies that we discuss in the succeeding sections, and should be made before considering estimation.
As a guide, we will use the notion of \emph{target trial emulation}, where we ``design'' an observational study as we would a randomized one, clearly defining the sample, including inclusion/exclusion criteria and exposure definition, outcome measurement, and the timing of all variables. For more on target trial emulation see \citet{Danaei2018,Dickerman2019} for a general discussion and \citet{benmichael_2021_trial} for a translation of these ideas to policy evaluations.
As we discuss below, these ideas are in contrast to traditional approaches for impact evaluation that rely heavily on parametric regression models, such as two-way fixed effects approaches, which do not have a clear design aspect but do have significant limitations and often lead to biased results \citep{goodmanbacon_did_2021}.

\subsection{Defining units and exposures}
\label{sec:exposures}
First, we must define the units in our study and the policy levels they are exposed to.
An initial step is to consider the jurisdictional level that the policy change of interest occurs at.
For example, RTC laws are state-level policy changes, and so it is appropriate to set our units of analysis to be states in the US.
Therefore, throughout this paper we will use ``units'' and ``states'' interchangeably to align with the running example of evaluating RTC laws.
However, our discussion is general to any unit of analysis.
Other policy changes occur at lower jurisdictional levels, e.g., at the county or municipality level, in which case we would set the unit of analysis to be at this lower level.
Throughout, we will say that we have $N$ states, indexed by $i=1,\ldots,N$.\footnote{As we will see below, we need not always set $N=51$ and include all states (plus Washington DC) in our analysis.}  We also assume that the data is available at that level, or that it has already been aggregated to that level.

Next, we must use a consistent definition of the policy levels, so that it is clear what we mean when we say that two states have the same ``treatment.''
In particular, in order for any policy impacts to be interpretable as
causal effects, there cannot be multiple versions of policies - or if there are, they should be characterized as different ``treatments.'' This, along with the no interference assumption discussed below, is often referred to as the Stable Unit Treatment Value Assumption (SUTVA) \citep{rubin_1980}.
Identifying what different versions of the policy of interest exist often requires legal research and qualitative methods to categorize policies; sometimes those who have conducted such research make their categorizations available as shared databases.

Except in special cases, even if two states have the ``same'' policy, the particulars of the implementation may differ greatly. This creates a practical challenge for policy evaluations. 
For example, returning to right-to-carry laws, historically there have been four different types of concealed carry weapons law: ``no-issue'', ``may-issue'', ``shall-issue'', and ``permitless'' carry laws. May-issue and shall-issue CCW laws create a permitting system that regulates who can legally carry a concealed weapon in public spaces, no-issue CCW laws do not offer a legal pathway for public citizens to carry a weapon in public, and permitless CCW laws do not require a permit to do so \citep{doucette_deregulation_2023, donohue_why_2023}.

Under no-issue CCW laws concealed carry is effectively banned.
Both may-issue and shall-issue CCW laws  provide a list of requirements that an applicant must satisfy to acquire a concealed carry permit, but may-issue laws allow law enforcement broad discretion over who is eligible to receive a CCW permit. In these states, potential permittees historically had to demonstrate ``proper cause,'' or demonstrate a need, for a CCW permit. The resulting regulatory regime makes obtaining a CCW permit difficult.
In contrast, shall-issue laws remove a potential permittee's need to demonstrate proper cause, likening the permitting system to a more administrative function where anyone who is eligible to own a gun under federal standards is also eligible to receive a CCW permit.
Finally, permitless CCW laws do not require any special vetting or licensing of individuals who wish to carry concealed firearms in public.
Because shall-issue and permitless laws make it significantly easier to carry a concealed gun in public spaces, these two policies are often combined under the ``right-to-carry'' title.
During the past 40 years, many states have removed restrictions on concealed gun carrying: in 2024 the vast majority of states have enacted some form of RTC law, while in 1980 the vast majority of states had not.\footnote{In 2022, the U.S. Supreme Court deemed the ``proper cause'' requirement unconstitutional in New York State Rifle \& Pistol Association v. Bruen. Though the eight remaining states with may-issue laws have largely maintained the remainder of their policies, these laws are under increased scrutiny in the wake of this decision and may not be as restrictive as they were in the past.
}

This leads to a trade-off between ``lumping'' policies that are broadly similar together and ``splitting'' policies apart based on their particulars.
The splitting approach allows for a more coherent definition of the treatment levels; however, this reduces the amount of data we have per treatment level and increases the number of potential comparisons combinatorially.
For example, if we were to separately consider shall-issue policies and permitless policies, we would estimate impacts for moving from (i) a no-issue to a shall-issue system; (ii) a may-issue to a shall-issue system; (iii) a may-issue to a permitless system; or (iv) a shall-issue to a permitless system.
The smaller sample sizes and larger number of comparisons would lead to reduced power, though the effects we estimate would be more interpretable.
As a result, researchers often ``lump'' policies together \citep{hasegawa_2020}.
In this analysis, we follow the original analysis of \citet{donohue2019right} and ``lump'' shall-issue and permitless systems together into a broad ``right-to-carry'' treatment, consistent with the typical framing of these laws.
As a result, any estimated effects can be thought of more properly as an average over the effects of shall-issue and permitless policies, a less interpretable measure but one with more statistical power to measure effects.

\subsection{Defining outcomes of interest}

Having defined the units (states) and the exposure levels (a broad notion of RTC), we move on to defining our outcomes of interest and their timing.
In general, most previous firearm policy evaluations have used either fatal firearm violence from the CDC's National Center for Vital Statistics (NCVS) or violent crime data from the FBI's  Uniform Crime Report (UCR). The CDC's NCVS provides counts of homicides, suicides, and unintentional deaths by mechanism (i.e., firearm or non-firearm) for all 50 states going back several decades using death certificate data. While this data is publicly available from the CDC's WISQAR's platform, that data is subject to censorship when the counts of a given state-year are below 10. This limits the usability of the publicly available dataset and necessitates that researchers apply for, obtain, and clean raw data provided by the NCVS. In order to obtain the raw NCVS data, one must (i) hold a terminal degree (e.g., PhD, ScD, MD etc.); (ii) articulate a clear research strategy for data use; and  (iii) agree to strict data storage and destruction requirements. However, because the raw NCVS data
rely on death certificates, there is a relatively low level of bias due to under-counting.

In contrast, the FBI's UCR is an uncensored, publicly available data set. However, it relies on self-reported data and likely contains moderate to high levels of bias. The UCR contains data reported by over 15,000+ local policy jurisdictions and provides data on offenses known to law enforcement (e.g., crimes reported to law enforcement), arrests made, and other criminal activities, such as the number of hate crimes or arsons \citep{banks_national_2016}. When using UCR data for policy evaluation it is important to consider the issue of missing data.
In particular, larger jurisdictions are more likely to  report data: the UCR relies on local law enforcement jurisdictions to self-report data, and larger jurisdictions have greater statistical capabilities to do so. Moreover, the FBI contacts jurisdictions representing populations of 100,000 or more to increase data reporting \citep{maltz_analysis_2006}. 
When using UCR data for state-level policy evaluation, it is important to give thought to whether and how the proportion of local jurisdictions that report data in a given state and year changes over time. A large fluctuation in reporting across states and years may be a result of large changes in the number of crimes or an artifact of changes in reporting practices.

With these caveats in mind, for our running example we will follow \citet{donohue2019right} and evaluate the impact of RTC on violent crime (defined as  murder, rape, robbery, and aggravated assault), using the FBI UCR data and dividing the number of violent crimes by state's population size.
We note that because of the potential issues with the UCR data, one should view this as a stylized analysis that elucidates the key methodological ideas.

One final consideration is the frequency with which outcomes are measured.
We will focus on \emph{yearly} crime measurements in our running example, so that we observe 37 crime rate measurements for each state from the years 1977-2014.
Finally, as a bit of notation, we will denote the violent crime rate per 100,000 residents for state $i$ in year $t$ as $Y_{it}$.

\subsection{Defining time zero}
\label{sec:time_zero}
\begin{figure}
  \centering
  \includegraphics[width = 0.6\textwidth]{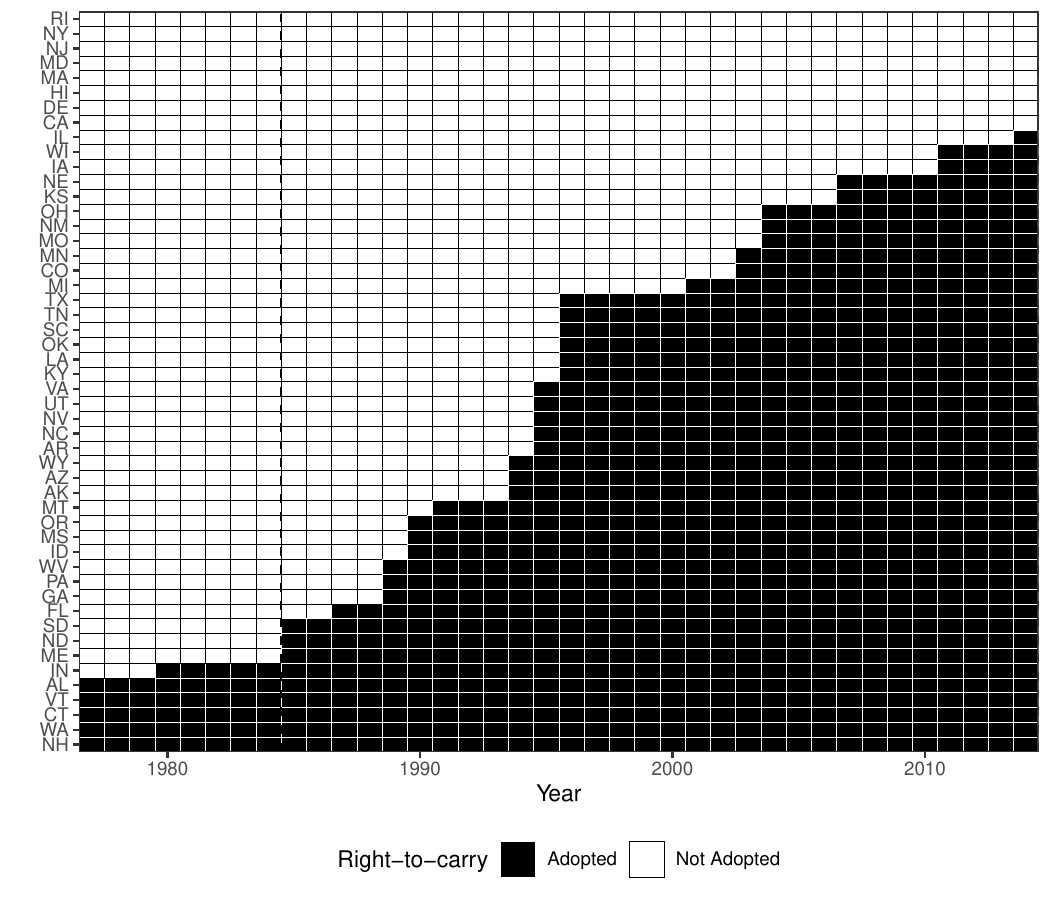}
  \caption{Timing of right-to-carry laws across the United States.}
  \label{fig:timing_plot}
\end{figure}

Next we must define a notion of ``time zero'' for each state: the year that a state may or may not have enacted an RTC law.
The key is that we must determine what are baseline measures pre-treatment and what are post-treatment outcome measures. This is easy for the states that did actually implement the policy but more challenging for the states that did not.
However, it is a crucial step as inadvertently conditioning on post-treatment values can lead to bias, even more than failing to adjust for potential confounders.

In public policy contexts, states can typically implement a policy change at any time.\footnote{A major exception are policy changes that are only possible due to other changes in law such as constitutional amendments or court decisions.}
For a treated state that eventually enacts an RTC law, we can define ``time zero'' as the year that the policy was enacted.
Formally, for a state $j$, we use $T_j$ to denote that year RTC was enacted, and we measure outcomes for a given ``event time'' $k$ centered around this year, $Y_{j T_j + k}$. Negative event times $k < 0$ denote pre-treatment measures, $k = 0$ denotes the year that RTC was enacted, and positive event times $k > 0$ denote post-treatment measures.
For example, Maine enacted RTC in $T_j = 1985$, so event time $k = -5$ corresponds to $T_j + k = 1980$, event time $k = 0$ corresponds to $T_j + k = 1985$, and event time $k = 5$ corresponds to $T_j + k = 1990$.

Figure~\ref{fig:timing_plot} shows what states enacted RTC in what year. There are several salient features.
First, we see in Figure~\ref{fig:timing_plot} the tell-tale ``staircase'' showing \emph{staggered adoption} of treatment: no state that enacted RTC repealed it during our panel from 1977-2014.
Second, we see that five states (NH, WA, CT, VT, and AL) had already enacted RTC before our first year of consistent data in 1977, and so pre-treatment measures with negative event times $k < 0$ do not exist.
As we will be using methods that require pre-treatment measures, we will have to remove these states from our sample.
In addition, we remove Indiana (RTC 1980) from our sample because there are only three years of pre-treatment data (i.e. $k = -3, -2, -1$), which is insufficient for identifying the pre-treatment trends that we will need to adjust for in later sections in order to reliably estimate effects.
Finally, eight states (CA, DE, HI, MD, MA, NJ, NY, and RI) did not enact RTC during the period 1977-2014. These are so-called ``never treated'' states, and we denote their treatment time as infinity, $T_j = \infty$.
These never treated states have no obviously defined time zero, but they will serve as comparators to states that enacted RTC during 1977-2014 in our analysis.

\subsection{Defining causal contrasts of interest}
\label{sec:causal_contrast}

With the various aspects of our dataset in place, we can now turn to defining our causal contrasts of interest.
To do so, we will use the potential outcomes framework \citep{neyman1923,rubin1974}, as adapted to this data setting by \citet{athey2018design}.
First, we need to create an expanded notion of the ``treatment.'' Recall that we have ``lumped'' the intervention into a broad RTC category so that there are two conditions for a state: an RTC law has been enacted, or it has not.
However, the timing of the policy change is also potentially important, and so we must track \emph{whether} and \emph{when} a state is treated in our analysis, using the treatment time variable $T_i$ defined above.

We will consider the potential violent crime rate for state $i$ in year $t$ if it were to enact RTC in year $s$, denoted as $Y_{it}(s)$.
In order to link these potential crime rates to the crime rates that we actually observe, we rely on the following two assumptions.
\begin{enumerate}
  \item \textbf{No interference between states}: a state is only affected by its own RTC status. If there are spillovers across states and this assumption is violated,  it becomes difficult to even define notions of causal effects. In practice while we might expect spillovers to exist, the no interference assumption may be a reasonable approximation.
  \item \textbf{No anticipation}: prior to enacting RTC, a state's potential violent crime rate is equal to its rate if it were never treated. We formalize this assumption as $Y_{it}(s) = Y_{it}(\infty)$ for $t < s$, with treatment time $s$. This assumption is sometimes highly implausible, for instance if the \emph{announcement} of a policy changes behavior before the policy itself takes effect.
  If such anticipation is plausible, we can re-define the treatment time to be an earlier date.
\end{enumerate}
Under these two assumptions we can link the observed outcomes to the potential outcomes as follows: if a state $i$ enacts RTC in year $T_i$, the outcomes for years prior to $T_i$ correspond to the ``never-treated'' potential outcome $Y_{it} = Y_{it}(\infty)$ and the outcomes for years $T_i$ and later correspond to the potential outcome when enacting an RTC law in year $T_i$, $Y_{it} = Y_{it}(T_i)$.

With all of this formalization, we can finally discuss what we mean by a ``causal effect.''
There are many potential comparisons and counterfactuals.
For example, we could focus on a single state, like Ohio (RCT 2004), and compare its violent crime rate in 2000 in a world where it had previously implemented RTC (say, in 1995) to a world where it had not yet implemented RTC.
These types of comparisons quickly become unwieldy.
A simple comparison often used in practice is to compare the observed crime rate in an RTC state to its counterfactual crime rate in a world where it never enacts RTC (formally, $Y_{it}(T_i) - Y_{it}(\infty)$).
To facilitate comparisons across states, we will turn to our discussion in Section~\ref{sec:time_zero} and measure this effect of RTC at event time $k$ years after adopting RTC.
Formally, we write this effect as $\tau_{ik} = Y_{iT_i + k}(T_i) - Y_{iT_i + k}(\infty)$.
Because we observe what happens under RTC ($Y_{iT_i + k}(T_i)$), the central methodological challenge is to impute what would have happened had the state not yet enacted RTC ($Y_{iT_i + k}(\infty)$).

While we are sometimes interested in estimating this state-level causal effect, it is difficult and noisy to estimate in practice.
Instead, we will focus on estimating an aggregate effect, the \emph{Average Treatment Effect on the Treated} states (ATT), $k$ years after each state adopted RTC: 
\[\text{ATT}_k = \frac{1}{N_\text{trt}}\sum_{i \mid T_i < \infty} \tau_{ik},\]
where $N_\text{trt}$ is the number of states that ever adopted RTC. This aggregate effect aligns the states according to event time, so that, e.g., $\text{ATT}_0$ corresponds to the average effect of RTC across states that enacted it, in the year that it was adopted, relative to a counterfactual scenario where the state had not adopted RTC by that year.

\section{Building blocks: estimating effects in comparative case studies}
\label{sec:single}
Only now that we have carefully considered \emph{what} policy effects we want to estimate, and how to interpret them, can we turn to \emph{how} to estimate them. In this section we will introduce the basic building blocks of two main approaches to estimating policy effects: differences in differences and synthetic controls. To do this, we will focus on using each estimator to try to measure the impact of RTC in a single state, which we can view as a comparative case study.
Although these state-level estimates are noisy, this will allow us to develop the key concepts for both approaches, which we will then show how to aggregate into a full analysis in Section~\ref{sec:aggregate}.
The two states we will use as running examples in the section are Ohio and West Virginia; each have aspects of their analysis that highlight important points for estimation. 

\subsection{Selecting units for comparison}

The first key choice in the analysis is carefully selecting comparison states for the focal state of interest. Our first concern is ensuring that we are making valid comparisons, comparing the focal state to other states that either never enact RTC, or enact RTC after the period of interest.
States that have already adopted RTC by the period of interest cannot serve as comparisons without strong (and typically unreasonable) modeling assumptions. With such a comparison set of ``donor states'', we can try to statistically adjust for any differences between the donors and the focal state. We will turn to this in the next section, but pause here to consider further design choices.

An important consideration in selecting comparison states is the length of the period for which we want to measure effects. Formally, we will say that we are interested in estimating effects for $K$ years after treatment.
We now have two options for constructing the donor set.
First, for each post-treatment year $k=0,1,\ldots,K$, we could construct a separate donor set consisting of all states that have not yet enacted RTC. This gives us the largest possible number of donors, but differences in impact estimates each year may be due to differences in the composition of the donor set, which would have to be carefully checked.
In this paper, we take an alternative approach: keeping the donor set the same for each treatment year by only including those states that do not enact RTC in the $K$ years after the focal state.

Formally, we define the set of possible donor units for treated state $j$ as those states $i$ which are treated at least $K+1$ periods after state $j$, which we denote as $\calD_{j}\equiv \{i: T_i \geq T_j+K + 1\}$. We will use $N_j$ to denote the number of donor units in $\calD_{j}$.
With this setup, the longer the post-period length $K$, the fewer donor states are available. If we choose a long enough post-period length, only never-treated states will serve as valid comparators, while setting a shorter post-period length allows states that are late adopters to serve as comparisons for early adopters.

In this paper, we will focus on effects up to $K=10$ years after a state enacts a right-to-carry law. For example, for West Virginia, which enacted RTC in 1989, we will measure effects until 1999; for Ohio (RTC in 2004) we will measure effects until 2014. For states that do not have $K=10$ years of post-RTC data, we will include as long of a post-period as possible. For example, for Wisconsin (RTC 2011) we will estimate effects for 3 years until 2014, and for Illinois, which enacted RTC in the final year of the panel (2014), we measure effects only for 2014.

\subsection{Difference-in-differences}
\label{sec:did}
\begin{figure}[t]
  \centering
  \begin{subfigure}{0.45\textwidth}
    \includegraphics[width=\maxwidth]{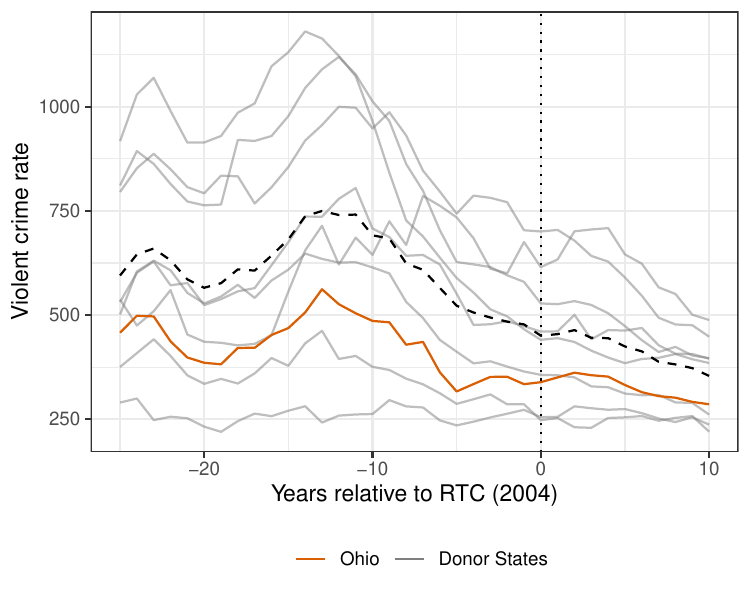}
    \caption{}
    \label{fig:oh_and_donors}
  \end{subfigure}
  \begin{subfigure}{0.45\textwidth}
    \includegraphics[width=\maxwidth]{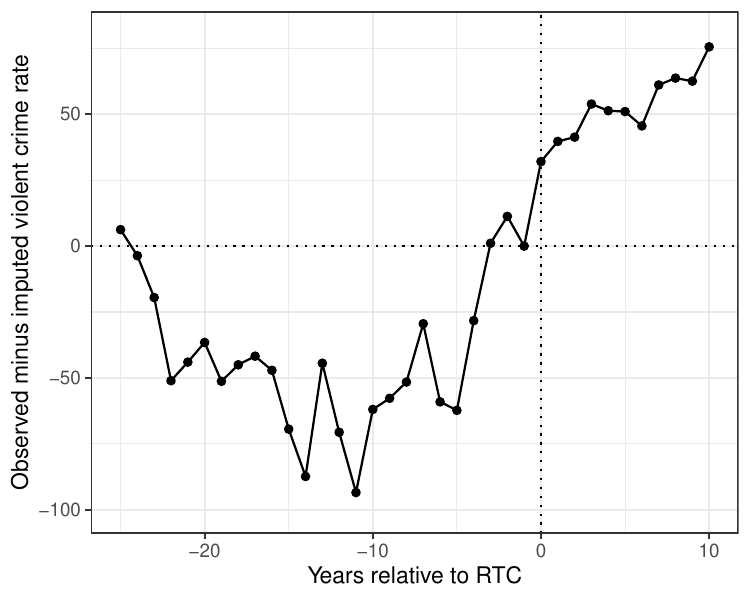}
    \caption{}
    \label{fig:did_case_study}
  \end{subfigure}
  
  \caption{(a) The violent crime rate for Ohio (orange) and donor states (grey) from 25 years before to 10 years after Ohio enacted a right-to-carry law. The dashed line represents the average violent crime rate among the donor states. (b) An ``event study plot'' of the difference-in-differences estimate of the impact of RTC on the violent crime rate in Ohio  pre- and post-RTC.}
  
\end{figure}

The workhorse method for estimating policy effects with panel data is the Difference-in-Differences method (DiD). The key idea of this approach is to compare the difference between the focal and comparison states before and after the policy change.
If there is a change in the difference after the policy is enacted, this could be evidence of an effect.
To build intuition for the DiD estimator, we will first consider two problematic comparisons, and discuss how DiD addresses the limitations of each.
Figure~\ref{fig:oh_and_donors} shows the violent crime rate for  Ohio and donor states from 25 years before and 10 years after Ohio enacted a right-to-carry law (2004), along with the average for the donor states.

The first comparison we can make is between the violent crime rates in Ohio and its comparison states in the ten years following Ohio enacting its RTC law in 2004. During these ten years, Ohio saw an average of 326 crimes per 100,000 residents, 21\% lower than the comparison states, which had an average of 417 crimes per 100,000.
Drawing causal conclusions from this simple comparison would ascribe all of the difference between Ohio and its comparison states to the RTC law, but Figure~\ref{fig:oh_and_donors} shows that in fact Ohio had a lower violent crime rate than its comparison states even before enacting its RTC law.

\begin{table}[t]
  \centering
  \begin{tabular}{l c c c}
    \toprule
  & \multicolumn{2}{c}{Right-to-Carry Law} &\\
  \cmidrule(lr){2-3}
   & Pre (1979-2003) & Post (2004 - 2010) & Difference\\ 
    \midrule
  Donors & 616.95 & 417.25 &  -199.7\\ 
  OH &  431.88 & 326.19 &  -105.69\\ 
  \midrule
  Difference & -185.07 & -91.06 & 94.01\\
     \bottomrule
  \end{tabular}
  \caption{A two-by-two differences in differences table for Ohio, showing the average violent crime rate per 100,000 residents in Ohio and its comparison states in the period before and after Ohio enated RTC.}
  \label{tab:2x2}
  \end{table}

As an alternative, we can instead inspect the violent crime rate in Ohio before and after enacting RTC. In the 25 years preceding the policy change, Ohio had an average of 432 violent crimes per 100,000 residents. In the 10 years afterwards there were an average of 326 violent crimes per 100,000 residents, a reduction of 16\%. However, drawing causal conclusions from this comparison requires us to believe that if Ohio did not enact an RTC law the violent crime rate would have remain unchanged from its historical average. Looking again at Figure~\ref{fig:oh_and_donors}, this seems implausible given that there was a reduction in the violent crime rate for all the donor states during the post-period as well.

The DiD estimator combines these two approaches by comparing the average difference between Ohio and its comparison states \emph{after} enacting RTC (91 fewer violent crimes per 100,000) to the difference \emph{before} enacting RTC (-185) to estimate an increase of 94 violent crimes per 100,000 in the post-RTC period (see Table~\ref{tab:2x2}).
We can view this as using the pre-RTC difference to adjust for pre-existing dissimilarities between Ohio and its donor states. 
An equivalent way to find this estimate is to compare the change in the average violent crime rate in Ohio before and after it enacted its RTC law in 2004 (-105.7), to the change in the violent crime rate in the donor states during the same period (-199.7).
We can view this characterization as using the pre-post difference in the donor states to adjust for any common changes in the violent crime rate across all states.

Under either view, after taking the difference of the differences, we are left with a component that may be ascribed to RTC. The key assumption underlying the validity of the DiD estimator is a \emph{parallel counterfactual trends} assumption.
This assumes that in the absence of Ohio enacting an RTC law in 2004, the trend in the violent crime rate would match the trend over that same time period in its comparison states \citep{card_minimum_2000,angrist2008mostly}.
Note that this assumption depends on our choice of scaling of the outcomes. For example, if there are parallel trends for the violent crime rate, then there typically will not be parallel trends for the \emph{log} violent crime rate or any other non-linear transformations without several stronger assumptions.

A threat to the parallel trends assumption is the  existence of \emph{time-varying confounding} factors. By comparing to the pre-RTC differences between the states, the DiD estimator can adjust for any fixed dissimilarities, even if they are not observed. However, if the states differ in ways that are not stable over time such as relative changes in demographic profiles or if the policy implementation process depends on features other than baseline violent crime rates --- e.g. changing policy in response to a particular spike in the crime rate in a particular year --- the parallel trends assumption will not hold.

The two-by-two comparison outlined here attempts to measure the impact of RTC averaged over the entire post-RTC period. However, we can adapt it to estimate effects for each individual post-RTC year.
To do so, we fix a common reference year, typically the year immediately preceding the policy change ($T_j - 1 = 2003$, in this case), then for every other year we compute the DiD estimator as above, using 2003 as the ``pre-period'' and using each other year as the ``post-period.''
Formally, we can write this estimator as 
\begin{equation}
  \label{eq:tau_jt_did}
    \hat{\tau}_{jk}^\text{did}  = Y_{j T_j+k} - Y_{j T_j - 1} -
                \frac{1}{N_j}\sum_{i \in \mathcal{D}_{j}}\left(Y_{i T_j + k} - Y_{i T_j - 1}\right).
\end{equation}
Figure~\ref{fig:did_case_study} shows these estimates in what is often called an ``event study'' plot. In addition, we can formalize the parallel counterfactual trends assumption as assuming that the expected change in the violent crime rates between the last pre-RTC period and every post-RTC period for treated state $j$ is equal to that for the donor states:\footnote{There are many potential ways to define this assumption. To keep our development concise we focus on this particular definition. See \citet{sun_did_2021,callaway_did_2021} for further discussion.}
\begin{equation}
  \E[Y_{jT_j + k}(\infty) - Y_{jT_j - 1}(\infty) ] = \E\left[\frac{1}{N_j}\sum_{i \in \calD_j}Y_{iT_j + k}(\infty) - Y_{iT_j - 1}(\infty) \right] \qquad \text{for all } k \geq 0.
\end{equation}

This more flexible estimation strategy allows us to construct a diagnostic measure for the parallel counterfactual trends assumption. The estimates $\hat{\tau}_{jk}^\text{did}$ for $k < 0$ measure the impact of RTC \emph{before} it is enacted. These serve as ``placebo'' estimates because they should be near zero; estimating non-zero impacts before the policy change occurs is a warning sign that at least one key assumption is violated.
Figure~\ref{fig:did_case_study} shows that there do not appear to be parallel trends, instead the event study plot is indicative of ``pre-trends'': about ten years before enacting RTC (i.e., in 1994) the violent crime rate in Ohio began to reduce more slowly than in its comparison states, so that we see a stark upward trend in the DiD estimates.
This indicates that the parallel trends assumption---or another key assumption such as the lack of anticipation---is likely to be violated and so we should not trust these estimates.

Note that this diagnostic is not a direct test of the parallel counterfactual trends assumption because the assumption is a statement about unobservable \emph{post-intervention} counterfactual trends.
However, it is a useful proxy: if there is evidence that trends are not parallel in the pre-RTC period then it is more difficult to argue that the counterfactual trends in the post period are parallel.\footnote{So you should not do it.}
Furthermore, as with any statistical test, lack of evidence against the parallel trends assumption is not evidence \emph{for} parallel trends \citep[see][for further discussion on equivalence tests for placebo checks]{Hartman2018}. One should not use noisy estimates around zero as a substitute for arguing why the assumptions hold.

\subsection{When parallel trends fails: synthetic controls}
\label{sec:scm}
The event study plot for Ohio in Figure~\ref{fig:did_case_study} precludes us from taking the DiD estimates seriously because there is evidence of differential trends in the pre-RTC period -- the violent crime rate in Ohio appears to have been on a different trajectory than the average of the other states.
To address this, we can try to find a \emph{weighted average} of comparison states, a \emph{synthetic control} state, that has a violent crime trajectory that resembles that in Ohio during the pre-RTC time period.

The synthetic control method (SCM), proposed by \citet{Abadie2003} and further developed in \citet{Abadie2010,Abadie2015} is an optimization-based procedure to try to find the best such weighted average. For a particular set of weights, we evaluate the quality of the resulting synthetic control by the squared difference between the violent crime rate in Ohio and the weighted average of the violent crime rates in the comparison states, averaged over the pre-RTC period. We can then find the optimal synthetic control that minimizes this average squared difference.

\begin{figure}[t]
  \centering
  \begin{subfigure}{0.45\textwidth}
    \includegraphics[width=\maxwidth]{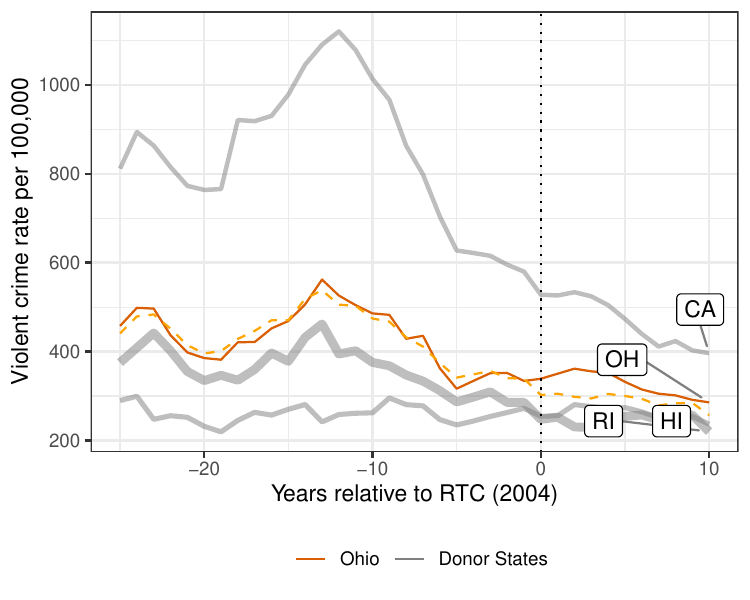}
    \caption{}
    \label{fig:oh_and_synth}
  \end{subfigure}
  \begin{subfigure}{0.45\textwidth}
    \includegraphics[width=\maxwidth]{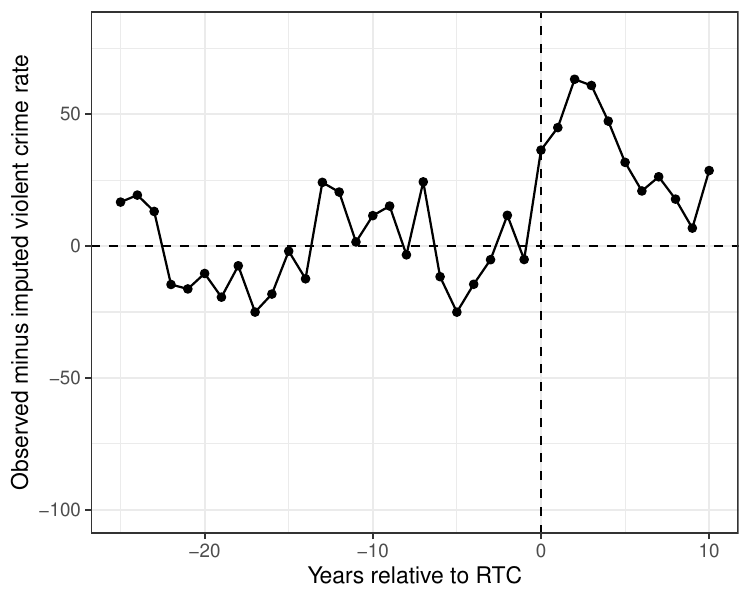}
    \caption{}
    \label{fig:scm_case_study}
  \end{subfigure}
  
  \caption{(a) The violent crime rate for Ohio (orange solid line) and donor states that form the synthetic control (grey, size proportional to weight), and the imputed violent crime rate for its synthetic control (orange dashed line) from 25 years before to 10 years after Ohio enacted a right-to-carry law. (b) A ``gap-plot'' showing the SCM estimates of the impact of RTC on the violent crime rate in Ohio pre- and post-RTC.}
  
\end{figure}

Formally, for treated state $j$, the SCM weights $\hat{\gamma}_j$ solve the following constrained optimization problem:
\begin{equation}
    \label{eq:scm_unit_j}
\begin{aligned}  
  \min_{\gamma_{1j},\ldots,\gamma_{Nj}} &  \;\; \frac{1}{L_j}\sum_{\ell = 1}^{L_j} \left(Y_{j T_j-\ell} \;-\; \sum_{i \in \mathcal{D}_j} \gamma_{ij} Y_{iT_j-\ell}\right)^2\\
  \text{subject to } & \sum_{i} \gamma_{ij}=1 \;\; \text{\&} \;\;  \gamma_{ij}\geq 0 \;\; \text{\&} \;\;  \gamma_{ij}=0 \text{ for } i \not \in \calD_{j}.
\end{aligned}
\end{equation}
The mathematical formalism of the constraints ensure that the weights (i) are non-negative ($\gamma_{ij}\geq 0$ for all $i$); (ii) sum to one ($\sum_{i} \gamma_{ij}=1$); and (iii) exclude impossible donors ($\gamma_{ij}=0$ whenever $i$ is not a possible donor, $i \not \in \calD_{j}$).
With these weights, the SCM estimate of the treatment effect compares the observed violent crime rate to the violent crime rate imputed by the synthetic control:
\begin{equation}
  \label{eq:scm_tau}
  \hat{\tau}^\text{scm}_{jk} = Y_{jT_j + k} - \sum_{i \in \mathcal{D}_j} \hat{\gamma}_{ij}\, Y_{iT_j + k}.
\end{equation}

A key feature of the SCM optimization problem \eqref{eq:scm_unit_j} is the constraint that the weights are non-negative and sum to one.
This constraint stops the synthetic control from \emph{extrapolating} away from the support of the donor units.
It has the additional effect that typically most of the SCM weights will be zero, and so the synthetic control will be a composite of only a few donor states, allowing for an interpretable comparison.

Returning to our running example in Ohio, Figure~\ref{fig:oh_and_synth} shows the violent crime rate per 100,000 residents in Ohio and its donor states, as in Figure~\ref{fig:oh_and_donors}, now with the donors' violent crime rate series sized proportional to their weight in Ohio's synthetic control. The main states that make up the synthetic control are  Rhode Island (60\%), Hawaii (20\%), and California (20\%); the other states contribute less than 1\% to the synthetic control. Taking the weighted average of these states gives the imputed violent crime rate from the synthetic control.

By finding an optimal combination of donor states, the synthetic control method allows us to weaken the assumption of parallel counterfactual trends on which the DiD estimator relies. SCM allows for the presence of unmeasured time-varying confounders; however, it requires that the relationship in the violent crime rate between states is stable over time.\footnote{Formally, SCM can estimate effects under a \emph{latent factor model} that imposes a particular structure on the relationship between the violent crime rate across states at the same time, and over time within the same state. For more technical discussion, interested readers should consult \citet{Abadie2010,Arkhangelsky2021_sdid,benmichael2021_ascm, benmichael2022_staggered}, among others.}
In this setting, SCM will provide good estimates of the treatment effect under two conditions: (i) the number of pre-treatment periods should be large and (ii) the pre-treatment fit --- the difference between the violent crime rate in Ohio and its synthetic control in the pre-RTC period --- must be good \citep{Abadie2010}.
Both conditions are important to ensure that the synthetic control is not biased: if the pre-period is short, then the synthetic control may be over-fitting to noise; if the pre-treatment fit is poor, then the synthetic control may be under-fitting to the signal.

While heuristics and rules of thumb for these two conditions are not readily available,
to evaluate the quality of the synthetic control we can use the same strategy as with the DiD estimator above: the SCM estimates of the impact of RTC in the pre-RTC period ($\hat{\tau}^\text{scm}_{jk}$ with $k <0$) should be near zero, and so they can again serve as placebo estimates. We can visualize this via an analog to the event-study plot in Figure~\ref{fig:did_case_study}: plotting $\hat{\tau}^\text{scm}_{jk}$ against the number of years before ($k < 0$) and after ($k \geq 0$) the implementation of RTC in Ohio. In synthetic control settings, this is often called a ``gap plot,'' shown in Figure~\ref{fig:scm_case_study}.\footnote{Note that the event study plot in Figure~\ref{fig:did_case_study} and the gap plot in Figure~\ref{fig:scm_case_study} are showing the same information, but for different estimators. In keeping with the nomenclature we will refer to these by separate names.}
We can see that the violent crime rate imputed by the synthetic control closely follows the observed violent crime rate in Ohio, but the fit is not perfect.
A useful one-number summary to diagnose the fit is the \emph{root mean square prediction error} (RMSPE), the square root of the average squared difference between the violent crime rate in Ohio and its synthetic control in the pre-period, which is equivalent to the square root of the objective in the SCM optimization problem \eqref{eq:scm_unit_j}. Using this metric, we see that the violent crime rate in Ohio differs from its synthetic control by roughly 17 violent crimes per 100,000 in the pre-period.

\subsection{When pre-treatment fit is poor: bias correction}
\label{sec:ascm}

\begin{figure}[t]
  \centering
  \begin{subfigure}{0.45\textwidth}
    \includegraphics[width=\maxwidth]{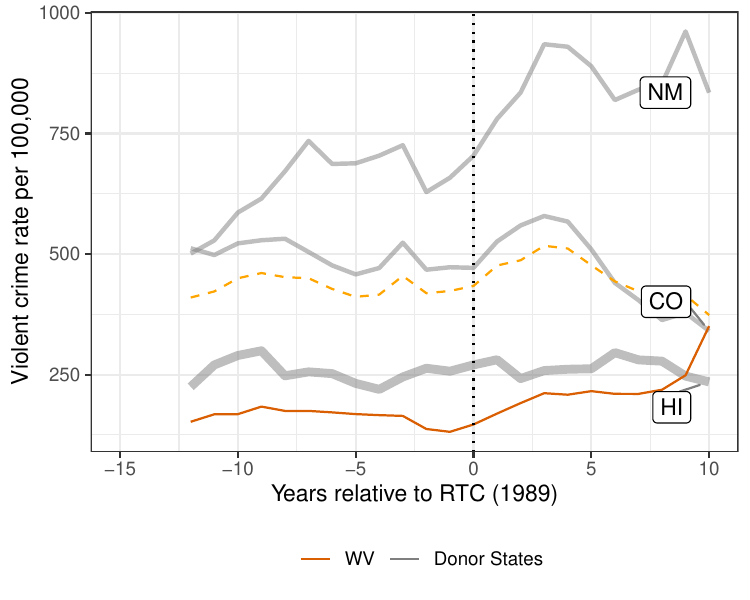}
    \caption{}
    \label{fig:wv_and_synth_int}
  \end{subfigure}
  \begin{subfigure}{0.45\textwidth}
    \includegraphics[width=\maxwidth]{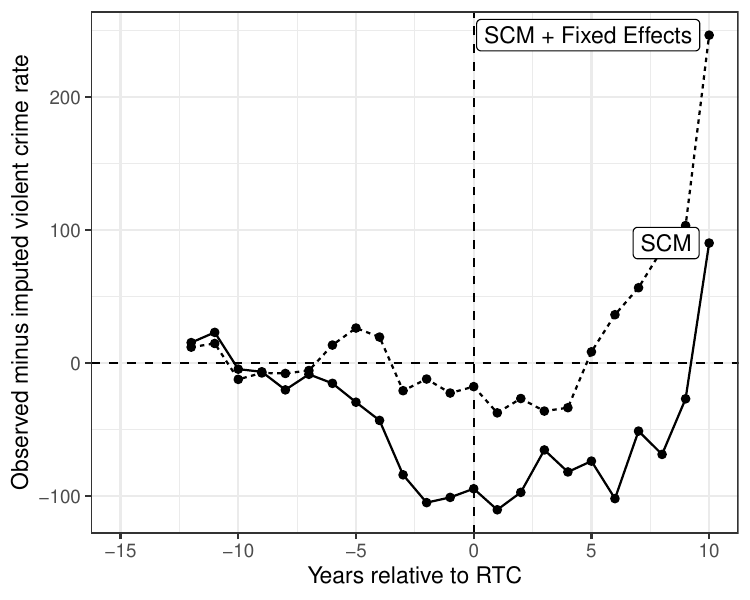}
    \caption{}
    \label{fig:scm_case_study_wv}
  \end{subfigure}
  
  \caption{(a) The violent crime rate for West Virginia (orange solid line) and donor states that form the synthetic control (grey, size proportional to weight), and the imputed violent crime rate for its synthetic control (orange dashed line) from 14 years before and 10 years after WV enacted a right-to-carry law.  (b) A ``gap-plot'' showing the SCM and fixed-effects augmented SCM estimates of the impact of RTC on the violent crime rate in WV pre- and post-RTC.}
  
\end{figure}

While the pre-treatment fit for Ohio is reasonable, it will not always be the case that it is possible to find a good synthetic control for a state. In fact, when studying the impact of a policy enacted in multiple states, it is likely that there is no good synthetic control for at least some of them. As an example of this, consider West Virginia (RTC in 1989). Following the procedure above gives a synthetic control for West Virginia comprised of Wisconsin (58\%) and Iowa (42\%), but this synthetic control is a poor fit. Figure~\ref{fig:scm_case_study_wv} shows the gap plot. The RMSPE is 52 violent crimes per 100,000, and the gap plot shows that in the years leading up to enacting RTC in 1989, WV began to differ substantially from its synthetic control. In 1988, immediately prior to enacting RTC, WV had 100 fewer violent crimes per 100,000 than its synthetic control. These estimates are not credible, but by construction we have found the best possible synthetic control. We now turn to what to do in these settings.

The lack of pre-treatment fit for West Virginia exhibited in Figure~\ref{fig:scm_case_study_wv} indicates that the synthetic control estimator is likely to be biased. To get a sense of how big this bias can be, we can attempt to directly estimate it. We can do this by first fitting a predictive model of the violent crime rate in the post-RTC period, if RTC were not enacted, using the donor states for WV. Then we can evaluate the difference between the predicted crime rate in WV and the weighted average of the predicted crime rates in the component states of its synthetic control (WI and IA). The difference between these two is the predictive model's prediction about the bias of the SCM estimate.
Formally, if $\widehat{Y}_{it}(\infty)$ is a prediction of the violent crime rate in the absence of RTC, then the estimate of the bias is
\[
  \widehat{\text{bias}}_{jk} = \widehat{Y}_{jT_j + k}(\infty) - \sum_{i \in \mathcal{D}_j} \hat{\gamma}_{ij}\, \widehat{Y}_{iT_j + k}(\infty).
\]
While this bias estimate can serve as a diagnostic tool, we can also use it to directly adjust the SCM estimate to \emph{correct} for the bias by simply subtracting out the estimate of the bias from the estimate of the treatment effect \citep{benmichael2021_ascm}. The bias-corrected (or augmented) synthetic control estimate is
\[
  \widehat{\tau}_{jk}^\text{ascm} = \widehat{\tau}^\text{scm}_{jk} - \widehat{\text{bias}}_{jk}.
\]

There are many potential predictive models to use to estimate and correct for the bias. \citet{benmichael2021_ascm} primarily advocate for using ridge regression to fit a predictive model of post-RTC crime rates on pre-RTC crime rates for states that did not enact RTC and then use this to predict the post-RTC crime rate in the absence of RTC for the treated states.
For simplicity, in this paper we will focus on another predictive model: two-way fixed effects estimation. This procedure estimates a state-level fixed effect $\hat{\phi}_i$ and a year-level fixed effect $\hat{\nu}_t$ to predict the crime rate for state $i$ in year $t$ in the absence of RTC as $\widehat{Y}_{it}(\infty) = \hat{\phi}_i + \hat{\nu}_t$. Two-way fixed effects models are intimately related to the DiD approach in Section~\ref{sec:did}, and in fact the fixed-effects augmented synthetic control estimate can be written as a form of a weighted DiD estimate:\footnote{The DiD estimate in Equation~\eqref{eq:tau_jt_did} uses the last pre-RTC period as the comparison point. In contrast, Equation~\eqref{eq:tau_jt_aug} uses all pre-RTC periods as comparison points and averages over these estimates.}
\begin{equation}
  \label{eq:tau_jt_aug}
    \hat{\tau}_{jk}^\text{fe-scm}  = \frac{1}{L_j}\sum_{\ell=1}^{L_j}\left[ \left(Y_{j T_j+k} - Y_{j T_j - \ell}\right) -
                \sum_{i \in \mathcal{D}_j}\hat{\gamma}_{ij}\left(Y_{i T_j + k} - Y_{i T_j - \ell}\right)\right].
\end{equation}
The fixed-effects augmented SCM has the same form as the DiD estimator \eqref{eq:tau_jt_did}, except that the pre-post differences in the violent crime rate for comparison units are weighted by the SCM weights. Conversely, setting the weights to be $1 / N_j$ above recovers the DiD estimate.

This estimator evaluates the bias by looking at the average difference in violent crime rates between WV and its synthetic control in the pre-RTC period. While this is a relatively simple estimate, it affords us several advantages.
Because this simple predictive model gives us estimates for all states in all years, we can fit the synthetic control on the \emph{residual} violent crime rate $Y_{it} - \hat{Y}_{it}(\infty)$; this is equivalent to modifying the SCM optimization problem to include a state-specific intercept \citep{Doudchenko2017, ferman2018revisiting, abadie2019synthreview}. This directly accounts for differences in the \emph{level} of the violent crime rate between states, allowing the synthetic control to focus on matching the \emph{trends} in the violent crime rate.
This is particularly helpful in contexts such as measuring impacts on crime rates, when the scale of the outcome varies across states and so it can be difficult to construct an adequate synthetic control that matches both the level and trends together.

Figure~\ref{fig:wv_and_synth_int} shows the violent crime rate in West Virginia, the three states that make up its fixed-effects augmented synthetic control (50\% Colorado; 35\% Hawaii; 14\% New Mexico), and the weighted average of the donor states before adjusting for level differences. The weighted average of these states has a much higher level of the violent crime rate than WV in the pre-RTC period (430 per 100,000 compared to 163 per 100,000 on average); however, the trend in the violent crime rate is a relatively good match.
By adjusting explicitly for level differences, the fixed-effects augmented synthetic control is free to use states such as New Mexico and Colorado that saw a slight decline in the violent crime rate during the late 1980's -- similar to West Virginia, but that had much higher levels.
As the gap plot in Figure~\ref{fig:scm_case_study_wv} shows, after adjusting for these level differences, the synthetic control has reasonably good fit, with an RMSPE of roughly 16 violent crimes per 100,000, much better than the un-augmented synthetic control, which has an RMSPE of 52 violent crimes per 100,000.

\paragraph{Additional estimation approaches.} A related approach to estimating effects when the synthetic control fit is poor is the ``Synthetic Differences in Differences'' approach proposed by \citet{Arkhangelsky2021_sdid}. This procedure estimates effects using a weighted DiD estimate with two types of weights: (i) weighting units by their synthetic control weight as in Equation~\eqref{eq:tau_jt_aug} above and (ii) weighting the pre-treatment time periods in the adjustment according to how the pre-treatment periods predict the post-treatment periods. Thus, we can view the estimate as a weighted form of the two-by-two DiD table (e.g. Table~\ref{tab:2x2}), using a pseudo-comparison group via SCM weighting and a pseudo-pre-treatment time via the time weights.
As with the augmented approach above, this can also be seen as a form of bias-correction using a model that predicts post-treatment outcomes in the absence of RTC from pre-treatment outcomes.

Finally, a broad class of estimation strategies directly model the outcomes under no treatment via an interactive fixed effects approach \citep[e.g.][]{Xu2018} or matrix completion \citep[e.g.][]{athey_matrix_2021}, or combine a low-rank matrix approximation with the synthetic controls objective \citep[e.g.][]{Amjad2018}. These approaches attempt to directly estimate and adjust for underlying latent confounders, taking advantage of modern matrix reconstruction techniques.
While these methods use different estimation approaches, the principles of the preceding sections are the same, including the design considerations in Section~\ref{sec:operationalize} and diagnostic checks such as the event study or gap plots.

\section{Aggregating effects under staggered adoption of policies}
\label{sec:aggregate}

Now that we understand the fundamentals of the DiD and SCM estimation strategies for a single state, we will turn to estimating aggregate effects across the treated states. 
Because we have laid the groundwork in the previous sections, these final steps will involve relatively simple and transparent aggregations.
Some methodological challenges arise due to aggregating estimates across time, but by aggregating across states we can get less noisy estimates and can construct confidence intervals for the effects via bootstrap re-sampling methods.

\subsection{From individual case studies to  time cohorts}

As a first step towards estimating aggregated effects, we will consider estimating effects for \emph{treatment time cohorts}, groups of states that enacted an RTC law in the same year. For example in 2004, the year Ohio enacted an RTC law, Missouri and New Mexico did as well; and Georgia and Pennsylvania enacted RTC laws in 1989 along with West Virginia. To estimate aggregate effects for the whole treatment time cohort with the DiD or (augmented) SCM approaches above, we can simply average the individual-level estimates.

Formally, we estimate the effect for states enacting RTC in year $s$, $k$ years after enactment as
\[
  \hat{\tau}_{sk} = \frac{1}{N_s} \sum_{j \mid T_j = s} \hat{\tau}_{jk},
\]
where $N_s$ is the number of states that enacted an RCT law in year $s$.
For DiD this is a straightforward average while for the SCM estimators this leads to another design choice: we can either find an individual synthetic control for each state, or find a single synthetic control for the \emph{average} violent crime rate in the treatment time cohort.
Typically, it is better to average across the time cohort: this adds precision over the noisy-state level measures and so the resulting synthetic control will typically both be able to achieve better pre-treatment fit and be less liable to over-fit to noise. For example, if we fit separate fixed-effects augmented synthetic controls for the three states in the 1989 cohort, the RMSPEs are 26 (GA), 3.6 (PA), and 16 (WV) violent crimes per 100,000.
In contrast, if we fit a single synthetic control for the average of the 1989 cohort, the RMSPE for the average is only 1.8 violent crimes per 100,000.

\subsection{Combining results across treatment time cohorts}

With these estimates aggregated to the time-cohort level in hand, we can follow the same idea to estimate the overall average treatment effect on the treated. For each year, we estimate the average treatment effect for that treatment-time cohort, then average across treatment time cohorts (weighting by the size of the cohort). We can write this estimator as
\[
  \widehat{\text{ATT}}_k = \frac{1}{N_\text{trt}}\sum_{s=1}^T N_s \hat{\tau}_{sk}.
\]

\begin{figure}[t]
  \centering
  \begin{subfigure}{0.45\textwidth}
    \includegraphics[width=\maxwidth]{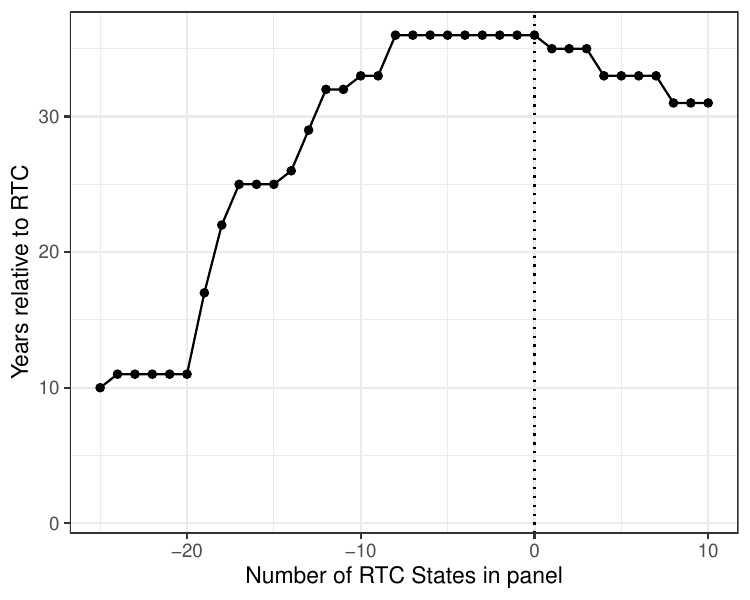}
    \caption{}
    \label{fig:num_treated}
  \end{subfigure}
  \begin{subfigure}{0.45\textwidth}
    \includegraphics[width=\maxwidth]{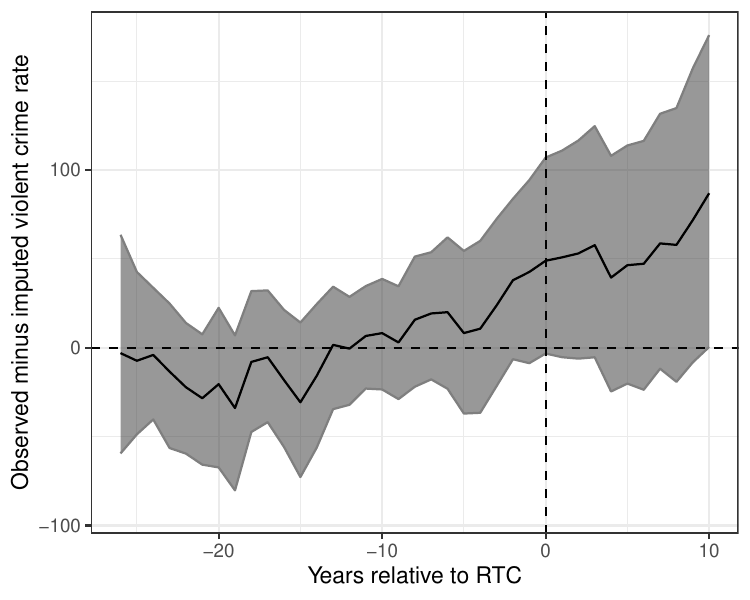}
    \caption{}
    \label{fig:event_study}
  \end{subfigure}
  
  \caption{(a) The number of states that ever enact RTC for which we observe crime rates versus the number of years prior to or after enacting RTC. (b) An aggregated ``event study plot'' of the difference-in-differences estimate of the impact of RTC on the violent crime rate, along with 95\% bootstrap confidence intervals.}
  
\end{figure}

Because we are aggregating across treatment-time cohorts, we must be careful about the differences between calendar time and event time (recall Section~\ref{sec:time_zero}). In particular, early adopters of RTC have fewer pre-RTC years from which to fit our estimators and late adopters have fewer post-RTC years from which to measure effects. So, we must restrict the estimate $\hat{\tau}_k$ to aggregate only over states for which we observe outcomes $k$ years post RTC.
For example, we can measure effects for all of the 36 different RTC states in the year that they enacted RTC ($k = 0$); however even for $k=1$ years post-RTC we lose Illinois because it enacted an RTC law at the end of our panel. By $k=10$ years post RTC the panel is reduced to 31 states (see Figure~\ref{fig:num_treated}).

\paragraph{Stacked DiD.}
When we take this approach with the DiD estimator, $\widehat{\text{ATT}}_k^\text{did}$ is a ``stacked'' difference in differences approach that aggregates and combines the individual or time-cohort level  two-by-two difference in differences tables as in Table~\ref{tab:2x2} in a transparent way that estimates a coherent causal effect \citep{cengiz_2019}.
This type of aggregated DiD estimator (proposed by \citet{callaway_did_2021} and \citet{sun_did_2021}, among others, with some particulars and details varying across approaches) allows for heterogeneity across treatment time cohorts and is the current recommended approach for estimating policy impacts with DiD.
This is in contrast to an older, regression-based approach that we recommend to avoid:  fitting a model for each state and year with an indicator for whether the state had enacted RTC by that year and fixed effects for state and year.\footnote{i.e. in \texttt{R} modeling notation, \texttt{outcome $\sim$ post\_rtc + state + year}. Note, though, that it is possible to reconstruct the DiD estimators we discuss via a regression with carefully constructed interaction terms \citep{wooldridge_twoway_2021}.} This approach has been shown to estimate a non-interpretable weighted combination of individual two-by-two group comparisons including some comparisons where both states are post-intervention \citep{goodmanbacon_did_2021}.
See \citet{sun_did_2021,callaway_did_2021} for further discussion and other forms of aggregation.

Figure~\ref{fig:event_study} shows the overall ``event study'' plot: our aggregate estimate $\hat{\tau}^\text{did}_k$ along with 95\% confidence intervals constructed via a boostrap re-sampling procedure (discussed below).
As in Section~\ref{sec:did}, we can use the estimates for years before the implementation ($k < 0$, left of the dashed lines) as a diagnostic check on the underlying parallel counterfactual trends assumption.
Here the parallel trends assumption does not seem plausible. In the years leading up to enacting RTC, RTC states saw violent crime rates rising relative to their comparison states, either because violent crime was rising at a faster rate, or because because they were decreasing at a slower rate. Therefore, even though the aggregate DiD estimates point to a weakly significant increase in violent crime due to RTC, these estimates are not credible.
Therefore, we will use augmented SCM to estimate the cohort-level effects and then aggregate. 

\paragraph{SCM with staggered adoption.}
As when aggregating from individual to time-cohort-level effect estimates with SCM estimators, there is a design choice on whether to find separate synthetic controls at the state or time-cohort level and then aggregate, or to find an aggregate synthetic control across states and time cohorts. Here there are methodological trade-offs due to the staggered adoption of RTC. Focusing on the average of the RTC states ensures that the violent crime rate for the  average synthetic control will match that of the average RTC state; however, because states enacted RTC in different years, averaging in this way opens us to potential \emph{interpolation} biases. If the underlying trends in the violent crime rate are changing substantially over time, averaging \emph{across} time-cohorts rather than \emph{within} them can lead to bias.
To circumvent this, \citet{benmichael2022_staggered} propose a ``partially pooled'' approach that simultaneously finds a synthetic control that fits well for the average state and within individual states or time cohorts. In this paper we will aggregate first to the time-cohort-level, then partially pool across time cohorts.

\begin{figure}[t]
  \centering
  \begin{subfigure}{0.45\textwidth}
    \includegraphics[width=\maxwidth]{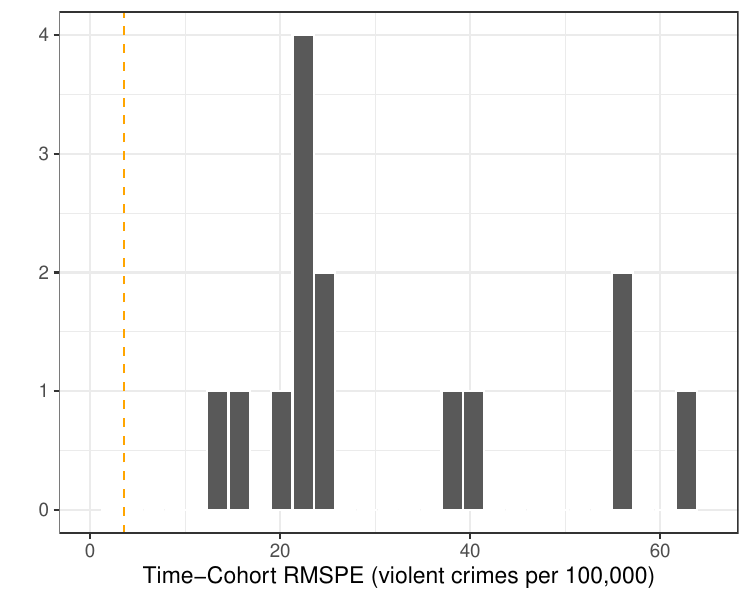}
    \caption{}
    \label{fig:rmspe_dist}
  \end{subfigure}
  \begin{subfigure}{0.45\textwidth}
    \includegraphics[width=\maxwidth]{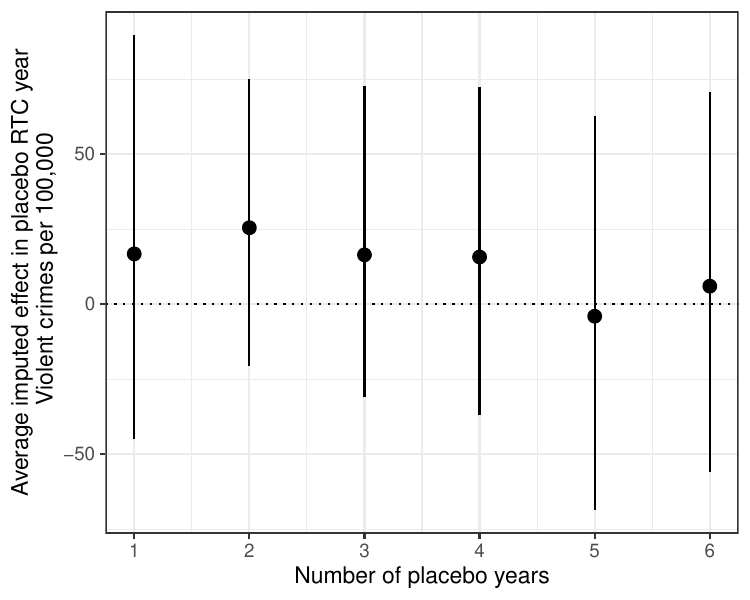}
    \caption{}
    \label{fig:placebo_in_time}
  \end{subfigure}
  
  \caption{(a) A histogram of the root mean square prediction error (RMSPE) for the fixed-effects augmented synthetic control for each time cohort. The dashed orange line represents the RMSPE for the aggregated synthetic control. (b) In-time placebo estimates and 95\% bootstrap confidence intervals of the impact of RTC in a held-out placebo RTC adoption year, setting the placebo adoption year to be 1-7 years before RTC was enacted in each state.}
  
\end{figure}

\begin{figure}[t]
  \centering
    \begin{subfigure}{0.45\textwidth}
      \includegraphics[width=\maxwidth]{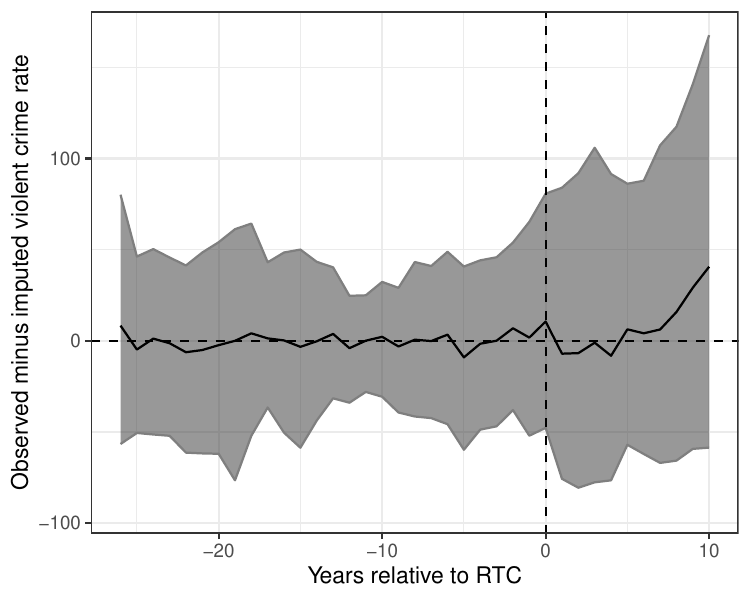}
      \caption{}
      \label{fig:fe_ascm_rate}
    \end{subfigure}
    \begin{subfigure}{0.45\textwidth}
      \includegraphics[width=\maxwidth]{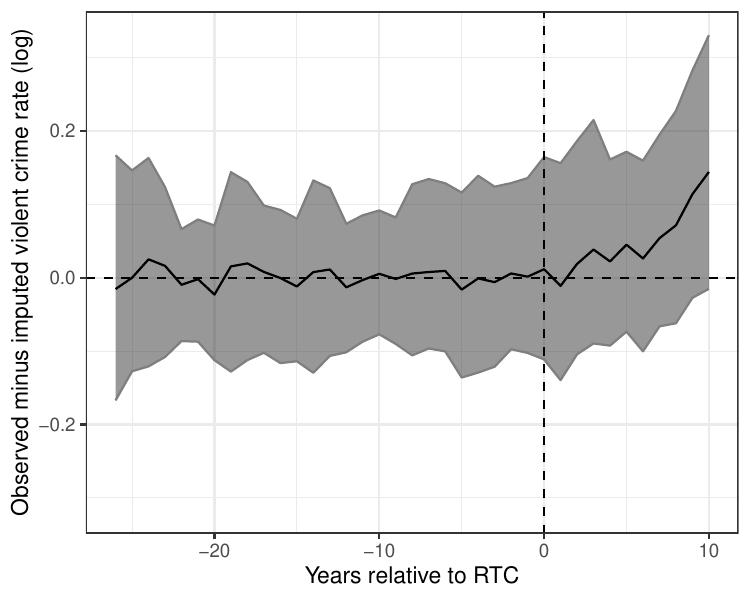}
      \caption{}
      \label{fig:fe_ascm_ln}
    \end{subfigure}
  
  \caption{An aggregate ``gap-plot'' showing the overal fixed-effects augmented SCM estimates of the impact of RTC on the violent crime rate in RTC states, along with 95\% confidence intervals for (a) the violent crime rate per 100,000 and (b) the log violent crime rate per 100,000.}
  \label{fig:fe_ascm}
  
\end{figure}

Figure~\ref{fig:rmspe_dist} shows the distribution of RMSPEs at the time-cohort level and for the average across time cohorts.
For most time-cohorts, the corresponding synthetic control fits reasonably well; the average time-cohort has an RMSPE of 31.6 crimes per 100,000, and the three time cohorts (1987, 1991, 2001) with RMSPEs above 40 crimes per 100,000 each consist of a single state (FL, MT, and MI, respectively).
The fixed-effects augmented synthetic control for the average across time cohorts fits very well, with an RMSPE of only 3.6 crimes per 100,000.
Figure~\ref{fig:fe_ascm_rate} shows the ``gap plot'' for the average time-cohort. As we expect from the RMSPE measure, the fit for the average across time cohorts is very good; on average the synthetic controls closely match their respective time cohorts so that the average difference is near zero during the entire pre-RTC period.

\paragraph{Approaches to inference.}
There are many possible approaches to inference with policy trial emulation. Our basic building block is to construct confidence intervals via re-sampling methods that re-sample entire states, such as the bootstrap, which are robust to correlations within a state over time.
The results presented in this paper use the \emph{wild boostrap} procedure proposed by \citet{Otsu2017} for matching estimators; see \citet{benmichael2022_staggered} for discussion on implementing this approach for panel data methods  and \citet{callaway_did_2021} for related bootstrap-style approaches relying on efficient influence functions.
Other approaches to inference are also available, including \emph{conformal inference} approaches that provide exact, non-asymptotic confidence intervals \citep{chernozhukov_exact_2021, Cattaneo2021, cattaneo_uncertainty_2023}.

When aggregating across multiple treatment time cohorts, an important consideration is that outcomes for the same state enter into the analysis in multiple places.
Therefore, a key step to constructing confidence intervals is to track where and how outcomes for the same state enter into the estimate, and then to use an inference approach that relies on re-sampling states (or directly computing standard errors).

\paragraph{Placebo checks and impact estimates.}
An important additional diagnostic is an out of sample check often called a \emph{placebo-in-time} check, wherein we set the RTC year for each state to be earlier than it actually is. For example, if we set the RTC year to be two years earlier, this gives us two years of pre-RTC data for each RTC state that are not used to fit the synthetic controls but where the estimated ``effect'' of RTC should be zero.
Figure~\ref{fig:placebo_in_time} shows the average estimated effect in the placebo RTC year, as we vary the placebo year from one to seven years before its actual value for each state.\footnote{Beyond seven years there would be no ``pre-treatment'' information for the first treatment time cohort.}
We find that while the placebo point estimates are not precisely zero, they average near 20 violent crimes per 100,000; zero is well within the range of 95\% confidence intervals, leading us to conclude that the synthetic control can reasonably impute the counterfactual violent crime rate.

These diagnostics and the event study plot in Figure~\ref{fig:event_study} lead us to the conclusion that the fixed-effects augmented SCM estimates will be more credible than the DiD estimates.
Figure~\ref{fig:fe_ascm_rate} shows the estimates for the impact of RTC laws on the violent crime rate per 100,000 (as measured using UCR data).
Overall, the point estimates are near zero for the first 5 years after enacting RTC, followed by an increase in the violent crime rate in the next 5 years. However, there is a substantial degree of uncertainty and the confidence intervals contain zero in every time period.

\paragraph{Estimating multiplicative effects with the log violent crime rate.}
A benefit of the approach laid out in this paper is that they key principles are agnostic to any transformations of the outcomes. So, for example, we can repeat the analysis above using the \emph{log violent crime rate per 100,000} as our outcome of interest. Estimating impacts on the log scale allows us to measure multiplicative effects, which approximate the percent change in the violent crime rate relative to the no RTC baseline.\footnote{Formally, the difference in log potential outcomes is the log ratio of the potential outcomes: $\log Y_{i T_i+k}(T_i) - \log Y_{i T_i + k}(\infty) = \log \frac{Y_{i T_i+k}(T_i)}{Y_{i T_i + k}} \approx \frac{Y_{i T_i+k}(T_i) -  Y_{i T_i + k}(\infty)}{ Y_{i T_i + k}(\infty)}$ for percent changes between roughly -20\% to 20\%.}
When evaluating impacts on transformed outcomes, one should conduct the same diagnostic and placebo checks as we have done above. In particular, for DiD the parallel trends assumption may seem plausible when evaluating the event study plot for some transformations of the outcomes, but not others. In the interest of space we do not report these diagnostics here -- they are similar to the ones above.
Figure~\ref{fig:fe_ascm_ln} shows the estimated impact of RTC laws on the log scale. When looking at multiplicative, rather than additive effects, we estimate that there was an increase in the violent crime rate over time, with an average increase of roughly 15\% 10 years after enacting RTC, albeit with confidence intervals that include zero.
The difference between the results with and without log-transforming the crime rate could be consistent with states with a lower baseline level of violent crime seeing a small absolute increase in violent crime that is still large relative to the baseline level.

This result for the log violent crime rate mirrors the results in the original analysis in \citet{donohue2019right}, which finds a 13-15\% increase in violent crime.
To contextualize these results, broadly researchers have found that RTC laws led to increases in some crime outcomes; for example,
\citet{doucette_deregulation_2023}  found that shall-issue laws were associated with roughly a 10\% increase in assaults with a gun. Other research has found the adoption of these laws increased the rate of officer-involved shootings of civilians \citep{doucette_officer-involved_2022} and has identified a link between shall-issue CCW law adoption and increases in firearm theft and decreases in homicide case closure rates \citep{donohue_why_2023}.

\section{Additional methodological challenges and future directions}
\label{sec:challenges}

In this paper we have considered the key components of operationalizing impact evaluations for gun policy research and estimating policy effects with modern statistical methods for causal inference. We have examined the key design decisions, assumptions underlying the analyses, and diagnostics for assessing these assumptions.
To conclude our discussion we will briefly touch on some of the additional methodological challenges that we have avoided in our development above because they constitute open problems in the field.

\paragraph{Estimating impacts on multiple outcomes.}
In our analysis we have focused on measuring the impact of RTC laws on violent crime, but this is an aggregate of several distinct types of crime. We could instead have considered the impact on  murder, rape, robbery, and aggravated assault separately, conducting a synthetic control analysis for each type of crime outcome.
There are two related challenges with such an approach.
First, finding synthetic controls separately for each crime outcome will lead to incompatible analyses.
For instance, in Figure~\ref{fig:wv_and_synth_int} we saw that the fixed-effects augmented synthetic control for West Virginia was composed of a combination of CO (50\%), HI (35\%), and NM (14\%), but if we instead analyzed the murder rate the augmented control would give a very different synthetic control, primarily weighting MA (36\%), MO (36\%), and HI (24\%).
If we were to analyze the four components of violent crime separately and then aggregate to estimate the impact on overall violent crime, we may get different estimates.
In addition, if we are interested in subgroup effects (e.g. homicides stratified by the race of the victim), we would also need to check or ensure that the sum of the estimated impacts on the subgroups, weighted by the population, is equal to the estimated impact overall.
Second, analyzing the outcomes separately precludes us from sharing information or structure across outcomes to get a better estimator.
There has been some methodological work to understand how to perform combined analyses across multiple outcomes simultaneously \citep[see, e.g.][]{Sun2023}, but optimally conducting such analyses remains an open question.

\paragraph{Measuring outcomes at different temporal resolutions.}
In our analyses we use violent crime measured annually; however, we could have used a finer temporal resolution, measuring crimes rates at the quarterly or monthly level.
Using quarterly or monthly crime rates would give us more pre-treatment time periods over which to fit our synthetic controls, which could lead to a better estimate. On the other hand, quarterly or monthly crime rates are noisier and can be subject to clear seasonal trends. \citet{sun_temporal_2024} show that the increase in noise often outweighs other benefits and a coarser aggregation can often be preferable. However, in some settings temporal aggregation can remove important signal in the data and lead to increased bias; the authors propose an approach that combines analyses at different temporal aggregations.
Understanding and diagnosing when this harmful signal removal can occur is an open methodological question.

\paragraph{Accounting for the structure of count (and often rare) outcomes.}
Most gun policy impact evaluations are focused on measuring the impact of the policy on \emph{counts}, whether they be counts of crimes, deaths, injuries, etc. In contrast, the theoretical development for synthetic control and related methods has primarily been focused on continuous outcomes, so rates (e.g., the violent crime rate per 100,000) are often analyzed as continuous. For large jurisdictions this approximation is typically adequate, but it can fail when jurisdictions are small. This is particularly a problem for analyzing the impact of county or municipality-level policy changes on violent crime and especially homicides: many smaller cities and towns see just a handful of homicides per year, if any. 
Extending the approaches described in this paper to properly account for such cases is another open direction for methodological work.

\paragraph{Accounting for multiple policies.}
As we discussed at the outset, there exist a wide variety of gun laws, each of which can have their own effect on gun violence. Moreover, different types of policies can \emph{interact} with each other, and certain combinations of policies may have a larger effect on gun violence than either alone. Formalizing such interactions
involves compound potential outcomes. For instance, to analyze the impact of RTC and stand-your-ground laws together we would construct a two-level potential outcome $Y_{it}(s, g)$ representing the violent crime rate in state $i$ in year $t$ if the state enacted an RTC law in year $s$ and a stand-your-ground law in year $g$; we could then consider counterfactual scenarios such as never enacting RTC but enacting a stand-your-ground law in a particular year.
Such comparisons very quickly become unwieldy.
However, without further (potentially quite strong) assumptions, it is not possible to avoid complex counterfactuals such as this.
An important area of methodological research is to understand whether there are reasonable scenarios where the complexity is limited enough to estimate compound policy impacts, and if there are, whether there is sufficient data to represent the different variations in combinations of tje policies, and then how to estimate those impacts.

\paragraph{Sensitivity analysis and a full policy evaluation workflow.}
In this paper we focused on designing and operationalizing a policy evaluation, and estimating impacts after doing so.
A final component to complete the policy evaluation ``workflow'' is to conduct a sensitivity analysis, inspecting how sensitive the results are to the underlying assumptions.
Formally conducting such sensitivity analyses involves characterizing how assumptions can be violated and then mapping how estimates would change if the assumptions were to be violated in those ways.
Approaches for doing so for the parallel trends assumption exist \citep[see, e.g.][]{roth_when_2023}, but developing such procedures for estimation strategies, such as SCM, that do not rely on the parallel trends assumption is an open area of methodological exploration.
Another form of sensitivity analysis is to consider sensitivity to data issues. For example, when using UCR data there may be differential under-reporting of data across states and time that could bias effects estimated under either the DiD or SCM approach.
Formalizing the potential for sensitivity to such bias is an open question.
Finally, flexible and robust approaches to statistical inference remain an active area of research.

\paragraph{Qualitative and mixed methods approaches.}
A final point relates to the role of qualitative and mixed methods research \citep[e.g.][]{mcginty_protocol_2021}.
When conducting policy evaluations there can be great value to using interviews and other qualitative or mixed methods approaches to gain information about what was happening ``on the ground'' in terms of the practical implementation of the policy and measurement and interpretation of the outcomes.
This can help to identify other contemporaneous policies or unique events that would cast doubt on the counterfactual parallel trends assumption.
It can also help to determine the true ``start date'' of a policy, especially if there is a gap between the legal enactment of a policy and its practical implementation.
 Such deep understanding can also help address data anomalies or strange patterns.  We encourage quantitative researchers to engage with substantive experts to ensure the appropriate use of data and interpretation of study results, leading to more useful and accurate inferences about policy effects.

\clearpage
\singlespacing
\bibliographystyle{apalike}
\bibliography{citations}

\end{document}